\documentclass[
reprint,
preprintnumbers,
nofootinbib,
superscriptaddress,
amsmath,amssymb,
aps,
prl,
]{revtex4-2}
\usepackage{bm}
\usepackage{CJKutf8}

\usepackage{graphicx, color}

\usepackage{microtype} 
\usepackage{dcolumn}
\usepackage[colorlinks=true,pdfstartview=FitV,linkcolor=blue,citecolor=magenta,urlcolor=blue,bookmarks=true,bookmarksnumbered=true]{hyperref}

\begin{document}

\preprint{RIKEN-iTHEMS-Report-26}
\preprint{HUPD-2604}

\title{Non-perturbative quark production and transport in the evolving glasma: \\the WAGASHI event generator}

\author{Nicholas~J.~Benoit}
\email{njbenoit@hiroshima-u.ac.jp}
\affiliation{Institute of Physics, Academia Sinica, Taipei, 11529, Taiwan}

\author{Chiho~Nonaka}
\email{nchiho@hiroshima-u.ac.jp}
\affiliation{Physics Program, Graduate School of Advanced Science and Engineering, Hiroshima University, Higashi-Hiroshima 739-8511, Japan}
\affiliation{Kobayashi Maskawa Institute, Nagoya University, Nagoya 464-8602, Japan}
\affiliation{International Institute for Sustainability with Knotted Chiral Meta Matter, Hiroshima University, Higashi-Hiroshima 739-8511, Japan}

\author{Hidetoshi~Taya}
\email{h\_taya@keio.jp}
\affiliation{Hiyoshi Department of Physics and Research and Education Center for Natural Sciences, Keio University, 4-1-1 Hiyoshi, Hiyoshi, Yokohama, Kanagawa 223-8521, Japan}
\affiliation{iTHEMS, TRIP Headquarters, RIKEN, Wako, Saitama 351-0198, Japan}

\date{\today}

\begin{abstract}
We study non-perturbative quark-antiquark pair production and the subsequent quark dynamics in the earliest glasma stage of relativistic heavy-ion collisions.
For that, we develop a new model, WAGASHI (Wong-precessing Anisotropic Glasma And ScHwinger-produced Initial-conditions), which combines classical Yang-Mills glasma evolution, Wong-equation transport, and Schwinger pair production on an event-by-event basis.  
We find that a sizeable number of quarks, comparable to the final hadron yields, are produced already during the glasma stage 
and subsequently undergo substantial momentum broadening and spin randomization, suggesting a significant contribution toward the early equilibration of the quark-gluon plasma.
We also determine the event-by-event distributions of baryon number, electric charge, strangeness, and spin polarization in Pb-Pb and O-O collisions at LHC energies, finding particularly large fluctuations in the smaller O-O system.
These results provide dynamical initial conditions for the subsequent hydrodynamic evolution of the quark-gluon plasma.
\end{abstract}

\maketitle

\paragraph{Introduction.---}
Over the past two decades, relativistic heavy-ion collision experiments at the Relativistic Heavy Ion Collider (RHIC) and the Large Hadron Collider (LHC) have firmly established the formation of the quark-gluon plasma (QGP) and revealed its remarkable properties~\cite{Yagi:2005yb, Busza:2018rrf}.
Despite this success, however, a major gap remains in our understanding of the collision dynamics: how the \textit{glasma}~\cite{Lappi:2006fp}, a far-from-equilibrium dense gluon state produced immediately after the collision, evolves toward the QGP is still poorly understood (cf. Refs.~\cite{Schlichting:2019abc, Berges:2020fwq} for the weak-coupling approach).
This uncertainty limits the quantitative extraction of QGP properties and obscures the interpretation of various observables that are sensitive to the early stage such as dilepton/photon production, anomalous transport, and spin polarization.
To overcome these limitations, settling the equilibration scenario is not enough; the initial distributions of conserved charges and spin must also be {\it quantitatively} calculated from the glasma stage.

The glasma is characterized by strong, coherent color electric and magnetic fields, whose typical magnitudes are $gE, gB \sim Q_s^2$, where $g$ is the coupling constant of the strong interaction and $Q_s$ is the saturation scale.
At the LHC energies, $Q_s$ is roughly $2 \sim 3$\;GeV and hence the color fields are far stronger than the mass scale of light quarks $m^2\ll gE, gB$~\cite{Gelfand:2016yho, Schlichting:2020wrv, McDonald:2023qwc, Matsuda:2024mmr}.
Such strong fields can excite quark-antiquark pairs directly from the vacuum by a {\it non-perturbative} mechanism, i.e., Schwinger pair production~\cite{Schwinger:1951nm} (Refs.~\cite{Gelis:2015kya,Fedotov:2022ely,Taya:2026llb} for review).

The Schwinger effect can serve as a natural early-stage quark-production mechanism from the glasma~\cite{Kajantie:1985jh, Gatoff:1987uf}.
Studies have focused on the pair production in flux-tube models~\cite{Bialas:1986mt, Bialas:1987en, Banerjee:1989by, tanji_dynamical_2009,tanji_quark_2010,tanji_electromagnetic_2015, Ruggieri:2015yea}, which connect the colliding ions with color-electric tubes.
Those tubes decay non-perturbatively to $\sim 250$ quarks before $\tau \lesssim 0.5$ fm/$c$~\cite{taya_quark_2017}.
Related calculations have treated quark-antiquark pair production from the glasma by numerically solving the Dirac equation, which also obtained the rapid quark production~\cite{gelis_quark-antiquark_2005, gelis_chemical_2006}.

Those non-perturbative mechanisms are complementary to the {\it perturbative} gluonic processes such as splitting $g\rightarrow q\bar{q}$ and fusion $gg\rightarrow q\bar{q}$~\cite{Kurkela:2018oqw, Du:2020zqg}.
The perturbative processes dominate in the dilute gluon regime, i.e., the late stage of the glasma evolution, at $\tau \gtrsim \alpha_s^{-3/2} Q_s^{-1} \approx 0.5\;{\rm fm}/c$ in the weak coupling limit ($\alpha_s := g^2/4\pi \approx 0.3$ and $\tau :=\sqrt{t^2 - z^2}$ is proper-time, where the $z$-axis is taken to be the beam direction)~\cite{Baier:2000sb, Schlichting:2019abc}.
Although early-time charge and flavor dynamics have been investigated within perturbative approaches~\cite{carzon_toward_2021} and Glauber-based models~\cite{giacalone_initial-state-driven_2025}, those dynamics have not been addressed by the previous Schwinger-effect studies.

To fully describe the quark transport consistently with the non-perturbative quark production from the evolving glasma, we develop a new code, WAGASHI (Wong-precessing Anisotropic Glasma And ScHwinger-produced Initial-conditions)\footnote{available for review at: \url{https://home.hiroshima-u.ac.jp/~njbenoit/wagashi/} later we plan to make it fully open-source}\footnote{WAGASHI ``\begin{CJK}{UTF8}{min}和菓子\end{CJK}" are traditional Japanese sweets.}.
The produced quarks are propagated through the glasma using the Wong equations~\cite{Wong:1970fu}, supplemented by spin-precession dynamics~\cite{Bargmann:1959gz}.
WAGASHI determines the time evolution not only of the phasespace distributions of the produced quarks but also of their charges and spin densities from the earliest stage of the collision.
It thereby provides a realistic quark initial condition for the subsequent hydrodynamic modeling of the QGP and a basis for assessing possible early-time imprints on quark fluctuations.

Development of such a code is important for studying the microscopic dynamics of quark production and transport in both large and small collision systems.
Particularly interesting are small collision systems, such as the recent O–O collisions~\cite{Brewer:2021kiv, CMS:2025tga, CMS:2025bta, ALICE:2026zck}.
In small systems, the event-by-event fluctuations are enhanced, making the charge and flavor distributions especially sensitive to the microscopic dynamics of quark production and transport.
Moreover, the shorter lifetime and reduced path length of the medium can limit rescatterings and reduce the washout of structures generated during the glasma stage.
O-O collisions may therefore offer a unique laboratory for isolating experimental signatures of quarks produced non-perturbatively at the earliest times.

\paragraph{Schwinger pair production in the glasma.---}
We implement the Schwinger pair production using the locally constant field approximation (LCFA)~\cite{Bulanov:2004de, Aleksandrov:2018zso, Taya:2020dco} together with an Abelian projection of the non-Abelian gauge fields.  Within LCFA, the fields in each spacetime cell are approximated as constant and homogeneous over the characteristic formation scale of the production process.  The local quark and antiquark production rates can then be evaluated using the exact constant-field result in a Lorentz frame in which the color electric and magnetic fields are parallel, ${\bm E}\parallel{\bm B}$~\cite{Nikishov:1969tt, Tanji:2010eu, Taya:2026llb}:
\begin{align}\label{eq:LCFASchwinger}
   &\frac{{\rm d}^7 N}{{\rm d}t{\rm d}^3 {\bm x} {\rm d}^3 {\bm p}} 
   = \sum_{\rm flavor} \sum_{i=1}^{N_{\rm c}=3} \sum_{s_\parallel=\pm \frac{1}{2}} \sum_{n=0}^\infty \frac{(g\omega_i)^2EB}{(2\pi)^3} {\rm e}^{-\pi\frac{m^2}{g\omega_iE}} \\
   &\times {\rm e}^{-\pi\frac{B}{E}(2n + 1 - 2s_\parallel)} \delta(p_\parallel)\delta^2\left( {\bm p}_\perp^2 - g\omega_iB(2n+1-2s_\parallel) \right) \;, \nonumber
\end{align}
where $n$ and $s_\parallel$ denote the Landau-level and spin quantum numbers, respectively, and ${\bm p}_\perp$ and $p_\parallel$ are the momenta perpendicular and parallel to the local fields, respectively.
Equation~(\ref{eq:LCFASchwinger}) is gauge-invariant after the summation over the quark-color index $i$~\cite{Nayak:2005pf}.
The weights $\omega_i$ control the effective couplings between the color field and the $i$-th colored quark and are determined by the algebra of the color group $SU(N_{\rm c})$~\cite{Gyulassy:1985oqt}.
The number of quarks, as well as antiquarks, produced in each spacetime cell is then sampled from a Poisson distribution whose mean is determined by Eq.~(\ref{eq:LCFASchwinger}).
Their spacetime positions and four-momenta are first generated in the parallel-field frame and are subsequently Lorentz-transformed back to the simulation frame.
Note that this local-rate implementation is substantially less expensive than a direct mode-function solution of the Dirac equation or the solving of the kinetic equation.
For a lattice with $N_{\rm site}$ sites in each spatial direction, evaluation of the local production rate scales as ${\mathcal O}(N^3_{\rm site})$, whereas a direct mode-function calculation scales nominally as ${\mathcal O}(N^6_{\rm site})$~\cite{gelis_chemical_2006, Gelis:2015kya}.
This reduction in computational cost is crucial for enabling event-by-event analyses.

Equation~(\ref{eq:LCFASchwinger}) relies on LCFA, which is valid for fields varying sufficiently slowly compared to the typical length scale for the Schwinger pair production and the size of the Landau-quantized wavefunction in a magnetic field~\cite{Aleksandrov:2018zso,Taya:2020dco,Fedotov:2022ely}; namely, $\Omega^{-1} \gg m/gE, m/gB$, where $\Omega^{-1}$ is the typical variation scale of the fields.
This condition is naturally satisfied for the glasma, for which $\Omega \sim Q_s$.

Additionally, Eq.~\eqref{eq:LCFASchwinger} relies on an Abelian approximation, in which the off-diagonal components of the gauge fields in the color space are neglected.
Such an Abelianization is exact for special backgrounds, including covariantly constant fields~\cite{Batalin:1976uv}, but is not generally justified for an arbitrary non-Abelian configuration~\cite{brown_vacuum_1979}.
To define the Abelian projection and suppress the off-diagonal components as much as possible, we fix the glasma fields to the maximally Abelian gauge (MAG)~\cite{schrock_coulomb_2013} and retain their Cartan components in evaluating the local production rate.

The color electromagnetic fields in Eq.~\eqref{eq:LCFASchwinger} are sourced by the glasma, determined by solving the classical Yang-Mills equation~\cite{avramescu_simulating_2023},
\begin{align} \label{eq:CYM}
    D_\mu F^{\mu\nu} = J^\nu \;,
\end{align}
where $F^{\mu\nu}$ is the non-Abelian field strength, $D_\mu$ is the covariant derivative, and $J^\mu$ is the source color-current carried by the colliding ions.
We solve Eq.~(\ref{eq:CYM}) on the real-time, boost-invariant, (2+1)-dimensional lattice~\cite{schenke_event-by-event_2012,muller_simulations_2020}.
We build a transverse profile from Monte Carlo sampling of a Woods-Saxon distribution for $^{208}$Pb and the PHOBOS generator for $^{16}$O~\cite{loizides_glauber_2026}.
The initial current $J^\mu$ is then determined from the McLerran-Venugopalan model~\cite{McLerran:1993ni, McLerran:1993ka, McLerran:1994vd}, whose parameters, $\mu$, were fit to the ALICE and CMS $0-5\%$ charged-particle multiplicities~\cite{alice_centrality_2025,collaboration_centrality_2026,adam_centrality_2016,aamodt_centrality_2011}.

\paragraph{Quark transport in the glasma.---}
After production, we use a colored particle-in-cell (cPIC) model to evolve the quarks and antiquarks in spacetime~\cite{muller_simulations_2020,avramescu_simulating_2023,dumitru_numerical_2006}.
That model solves the Wong equations for the phasespace $(x^\mu, p^\mu, {\mathcal Q})$~\cite{Wong:1970fu}, together with the colored extension of the Bargmann-Michel-Telegdi (BMT) equation for the spin four-vector $S^\mu$~\cite{Bargmann:1959gz, elze_quark-gluon_1989, Yang:2021fea, Hidaka:2022dmn}:
\begin{align}
\begin{split}
    &\frac{{\rm d} x^{\mu}}{{\rm d}t} = \frac{p^{\mu}}{p^0}\;, \ 
    \frac{{\rm d} {\mathcal Q}}{{\rm d}t} = {\rm i}g \frac{p^{\mu}}{p^0} [A_{\mu}, {\mathcal Q}] \;, \label{eq:wongBMT} \\
    &\frac{{\rm d} S^{\mu}}{{\rm d}t}  = 2 g \,{\rm Tr}\,[ {\mathcal Q} F^{\mu \nu} ] \frac{S_{\nu}}{p^0} \;, \ 
    \frac{{\rm d} p^{\mu}}{{\rm d}t}  = 2 g \,{\rm Tr}\,[ {\mathcal Q} F^{\mu \nu}] \frac{p_{\nu}}{p^0} \;,  
\end{split}
\end{align}
where ${\mathcal Q}$ is the color charge and ${\rm Tr}$ denotes the trace in the color space.

\begin{figure}[hbt]
    \includegraphics[width=\linewidth]{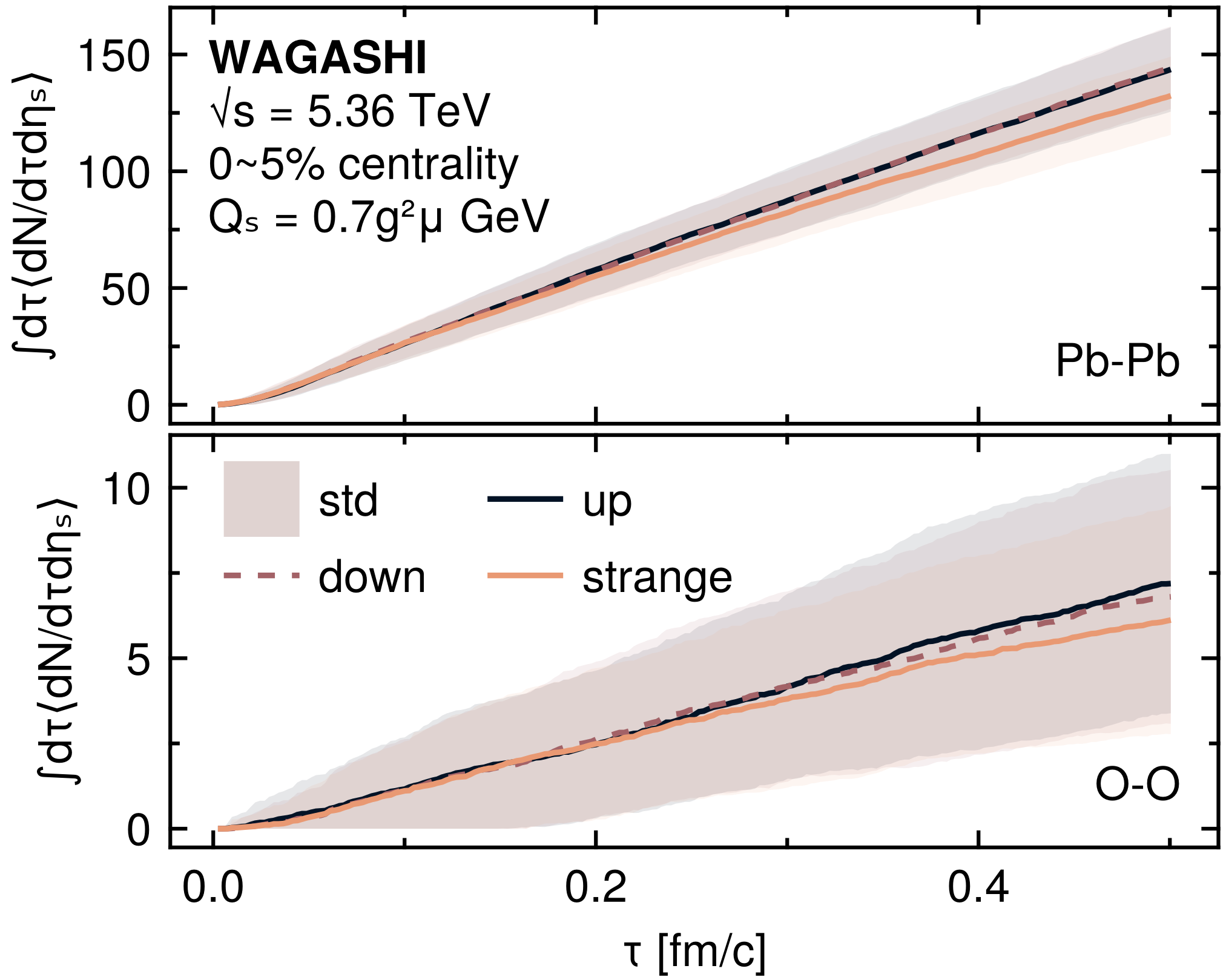}
    \caption{
       The total number of quarks and antiquarks produced per space rapidity $\eta_s$ in $0-5\%$ centrality by WAGASHI.
       The lines are the event-averaged (200 events) mean, while the shaded bands indicate the corresponding standard deviation (``std") arising from event-by-event fluctuations.
    }
    \label{fig:quarknumber}
\end{figure}
\paragraph{Time evolution of quark production.---}
We find that a sizeable number of quarks is produced via the Schwinger effect, Eq.\eqref{eq:LCFASchwinger}, already at early times, before the perturbative processes become effective, reaching $dN/d\eta_s \approx 400$ quarks and antiquarks at $\tau \sim \alpha_s^{-3/2}Q_s^{-1} \approx 0.5\;{\rm fm}/c$.
Figure~\ref{fig:quarknumber} shows the number of produced quarks increases approximately linearly with time, although the strange-quark production rate exhibits a slight slowdown at $\tau \gtrsim Q_s^{-1} \approx 0.1\;{\rm fm}/c$.
This behavior occurs because the glasma decays as the system expands longitudinally.
As the field strength decreases, the mass-dependent exponential suppression of the Schwinger effect in Eq.~(\ref{eq:LCFASchwinger}) becomes relevant for strange quarks, while remaining negligible for the much lighter up and down quarks.
Nonetheless, this slowdown is not significant over the time scale $\tau \lesssim \alpha_s^{-3/2}Q_s^{-1}$, during which the coherent-glasma description is applicable, and the strange-quark yield remains about $90\%$ of the up-quark yield.

We find a similar flavor composition in O-O collisions, although the total number of produced quarks is much smaller $\approx 15$.
Since the nuclear radius scales as $R \propto A^{1/3}$ and the saturation scale as $Q_s^2 \propto A^{1/3}$, the yield at fixed $\tau$ from Eq.~(\ref{eq:LCFASchwinger}) scales as $N \propto R^2 Q_s^4 \propto A^{4/3}$.  This gives $N_{\rm Pb\mathchar`-Pb}/N_{\rm O\mathchar`-O} \approx (208/16)^{4/3} \approx 30$, consistent with the observed yield difference.

\begin{figure}[hbt]
    \includegraphics[width=\linewidth]{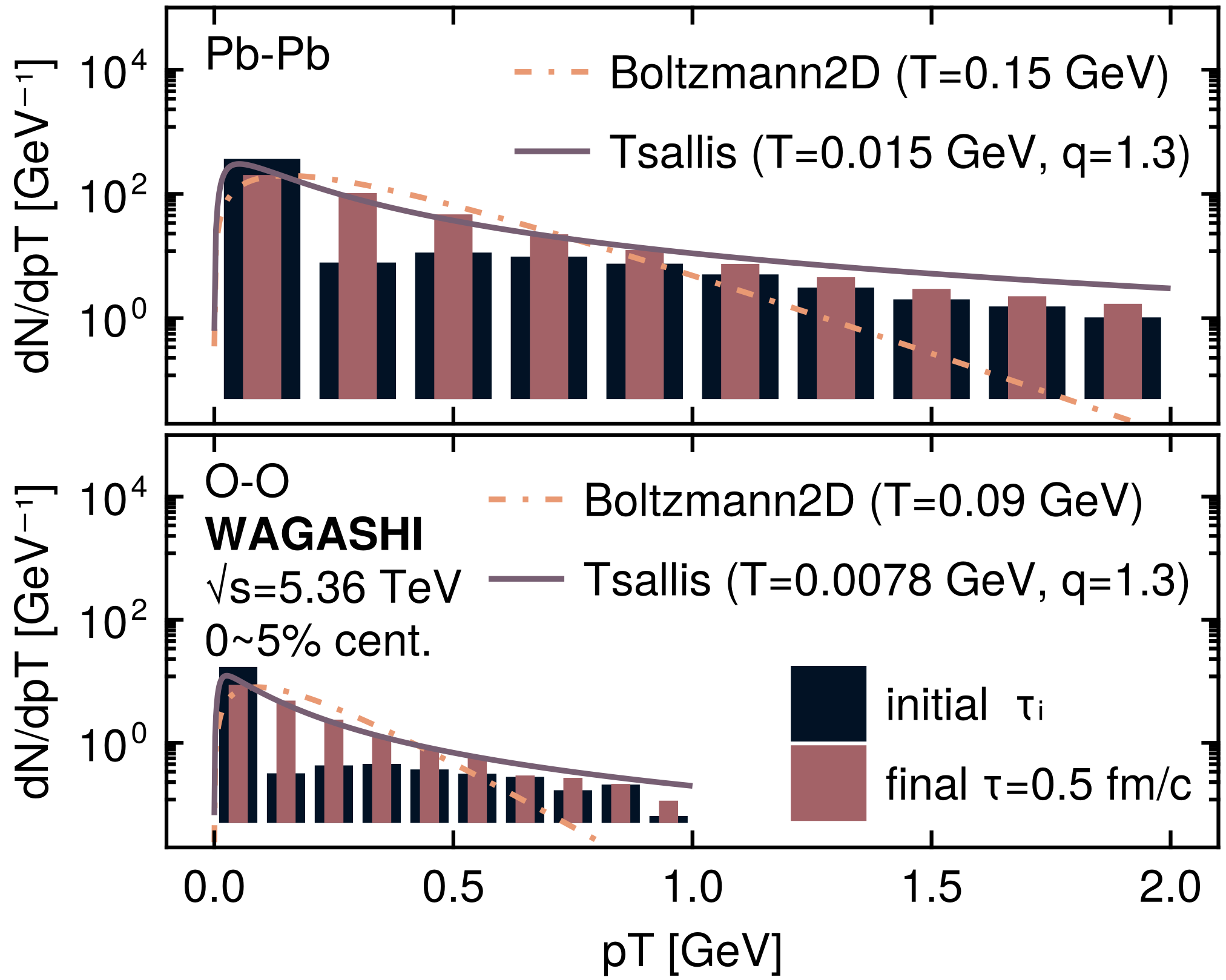}
    \caption{
       Transverse momentum spectrum of the produced quarks (200 event averaged), together with thermal Boltzmann $\propto p_{\rm T} {\rm exp}[-p_{\rm T}/T]$ and Tsallis $\propto p_{\rm T} [1 + (1-q) p_{\rm T}/T_q]^{q/(1-q)}$ distributions. 
    }
    \label{fig:quarkmom}
\end{figure}

Figure~\ref{fig:quarkmom} shows the transverse-momentum spectrum of the produced quarks at their production times $\tau_i$ and at a fixed proper-time $\tau = 0.5\;{\rm fm}/c$.
The initial spectrum is concentrated at very low transverse momenta, $p_{\rm T} \sim 0$, as a consequence of the Landau quantization by the strong color magnetic field.
As the quarks propagate through the glasma, they gain momentum through the color Lorentz force~\eqref{eq:wongBMT} and undergo a random walk in momentum space, reflecting the spatially varying and effectively random orientations of the glasma.
Consequently, the distribution broadens toward larger $p_{\rm T}$.
This momentum broadening represents a transfer of the produced quarks from soft to hard momentum modes, contrasting with the hard-to-soft energy flow characteristic of perturbative bottom-up thermalization~\cite{Cabodevila:2023htm, barrera_cabodevila_quark_2025}.

We find that the spectrum at low transverse momenta, $p_{\rm T} \lesssim 1\;{\rm GeV}$, exhibits an approximately thermal shape, as indicated by the fit to a Boltzmann distribution in Fig.~\ref{fig:quarkmom}, with an effective temperature of $T \approx 150\;{\rm MeV}$.
This suggests that the early glasma stage, $\tau \lesssim \alpha_s^{-3/2}Q_s^{-1}$, may facilitate subsequent kinetic equilibration by preconditioning the quark momentum spectrum.
We emphasize that this thermal-like spectral shape at low $p_{\rm T}$ does not by itself establish kinetic equilibration of the full quark sector.
The glasma remains a coherent field-dominated system and lacks the incoherent quark-gluon scatterings that occur at later stages of the evolution.
These are reflected in the clear deviation from the Boltzmann fit at $p_{\rm T} \gtrsim 1\;{\rm GeV}$.
Indeed, the spectrum over the full $p_{\rm T}$ range is better described by a Tsallis distribution~\cite{Tsallis:1987eu}, with the non-thermal parameter $q$ deviating from unity.

\paragraph{Fluctuations of conserved charges.---}
\begin{figure*}[thb]
   \includegraphics[width=\linewidth]{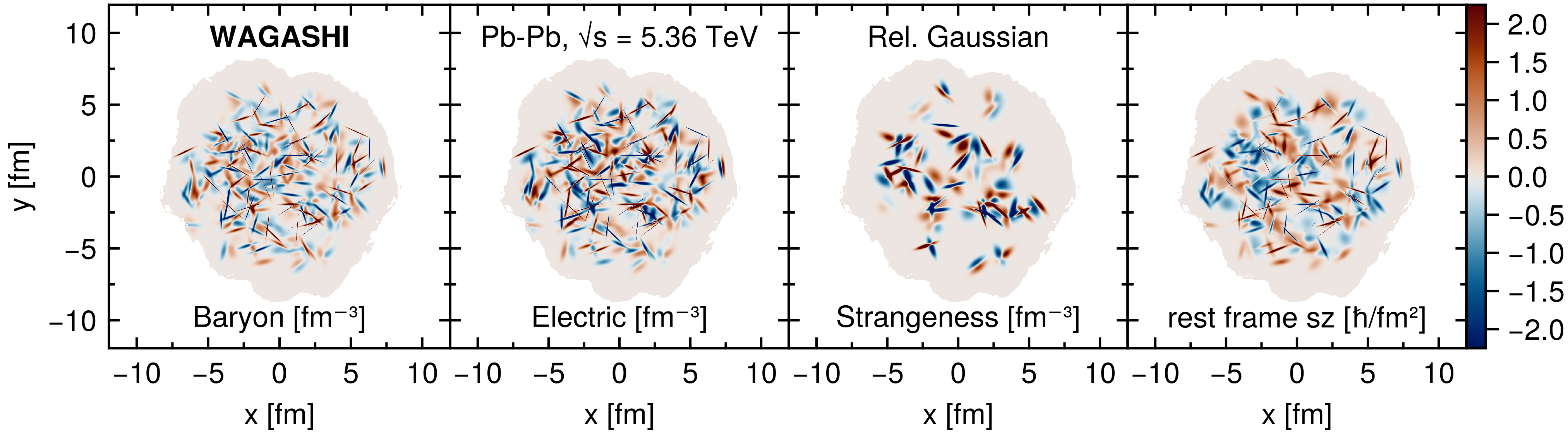}
   \caption{The conserved charge $n_q$ and spin $s_z$ densities in a single Pb-Pb event, obtained using WAGASHI.  
   Relativistic Gaussian smearing contracts the deposited density in the direction of transverse motion, which creates the line-like structures.
   }
   \label{fig:chargedistro}
\end{figure*}
The conserved charge densities are highly fluctuating in the transverse plane, reflecting the random field profile of the glasma and the stochastic nature of the Schwinger pair production.
This is shown in Fig.~\ref{fig:chargedistro}, where we calculated the density profiles of a single WAGASHI event by using the relativistic Gaussian smearing, with a smearing width $0.4\;{\rm fm}$ chosen for a demonstration.
That smearing contracts the charge density along the quarks' transverse momenta, creating the line-like structures.
The abundance of those structures suggests that there is a large transverse separation between quark-antiquark pairs due to the Wong evolution~\eqref{eq:wongBMT}; otherwise, pair contributions would self-cancel.
Studies indicate initial fluctuations like ours can survive the hydrodynamic evolution.
They have estimated a $5 \sim 10$ reduction for ideal hydrodynamics~\cite{plumberg_bsq_2024,carzon_toward_2021}, which would reduce our peak densities to $0.4 \sim 0.2\text{ fm}^{-3}$ but are still sizeable.

\begin{figure}[htb]
   \includegraphics[width=\linewidth]{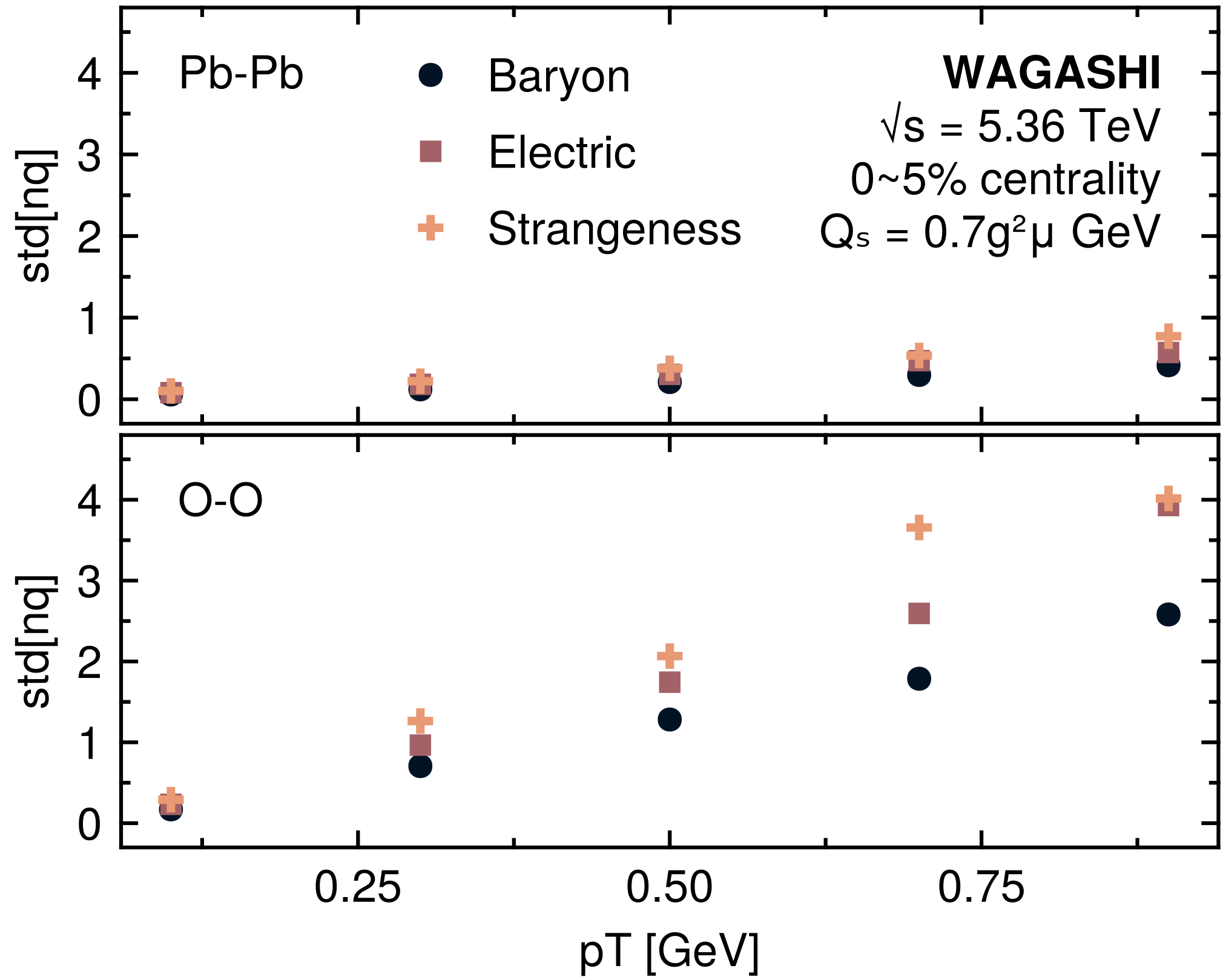}
   \caption{
        Momentum $p_{\rm T}$ dependence of fluctuations of the quark and antiquark charges.  
   }
   \label{fig:chargeflux}
\end{figure}

Those fluctuations appear not only in coordinate space but also in momentum space.
See Fig.~\ref{fig:chargeflux}, where we sort the charges $n_q$ per $p_{\rm T}$ bin and take the event-by-event standard deviation normalized by the mean number of quarks and antiquarks in the bin $\langle N(p_{\rm T})\rangle$,
\begin{equation}
    \text{std}[n_q(p_{\rm T})] := \frac{\sqrt{\langle n_q^2(p_{\rm T}) \rangle}}{\langle N(p_{\rm T})\rangle} \;.
\end{equation}
The fluctuations become larger at higher $p_{\rm T}$ and for smaller systems, where the number of quarks and antiquarks in each $p_{\rm T}$ bin is smaller.
The charge fluctuations grow during the Wong evolution in the glasma as quarks and antiquarks acquire different momenta, changing $p_{\rm T}$ bins.
Those fluctuations can appear in charge-dependent observables.
For example, the $p_{\rm T}$-dependent flow 
can be different for hadrons of opposite charges.

\begin{figure}[htb]
   \includegraphics[width=\linewidth]{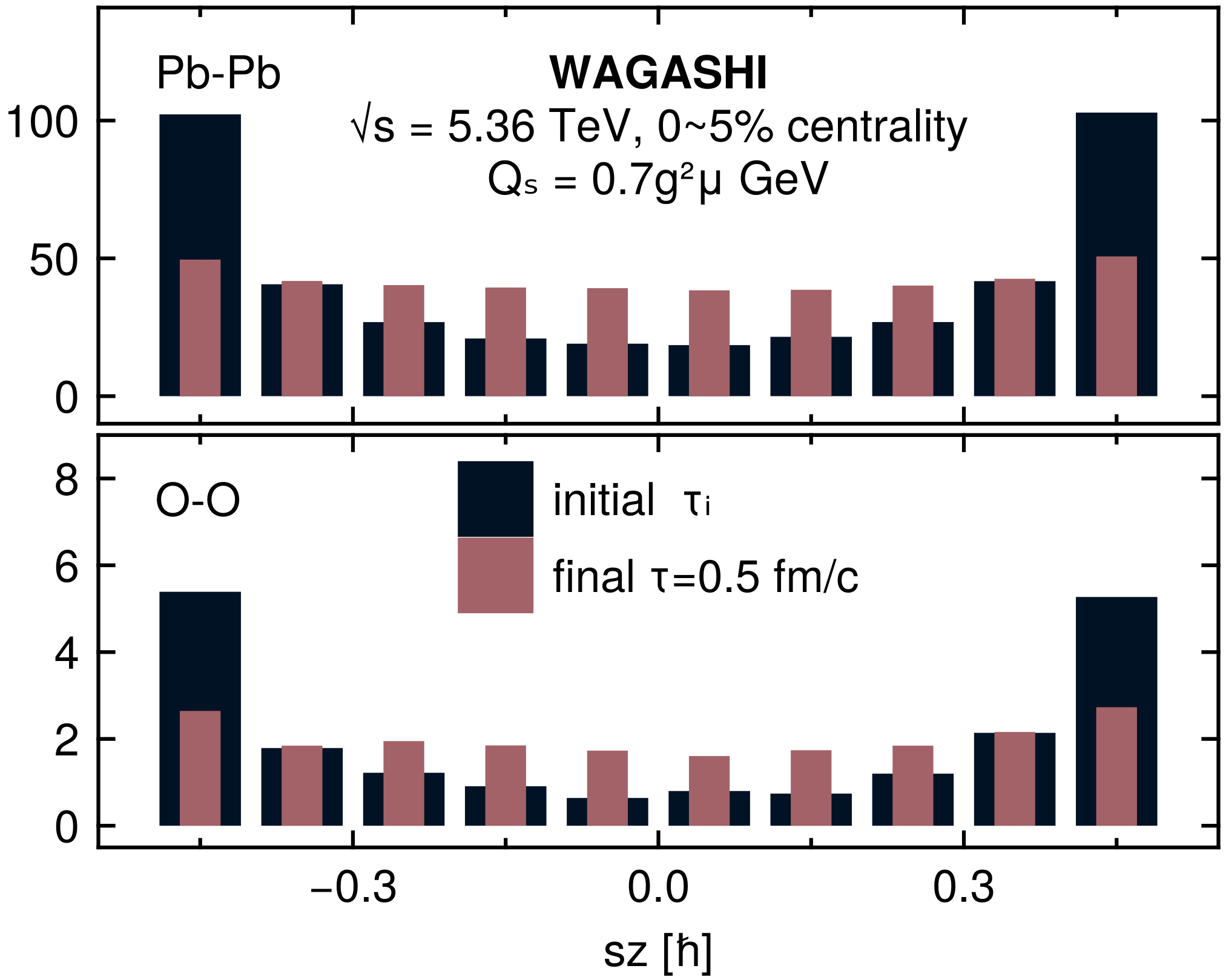}
   \caption{
        Binned rest frame spin $s_z$ at the production time and final simulation time $\tau=0.5\;{\rm fm}/c$ from WAGASHI.
   }
   \label{fig:restspin}
\end{figure}

\paragraph{Spin-polarization.---}
Finally, we discuss the spin polarization of quarks.  The quark spins are strongly polarized along the beam direction at the time of production as a consequence of Landau quantization in the strong color magnetic field; see Fig.~\ref{fig:restspin}.
As the quarks subsequently propagate through the glasma, their spins undergo stochastic precession through the colored-BMT evolution~(\ref{eq:wongBMT}), resulting in a progressive randomization analogous to the momentum broadening shown in Fig.~\ref{fig:quarkmom}.
By $\tau \sim \alpha_s^{-3/2}Q_s^{-1} \approx 0.5\;{\rm fm}/c$, the spin orientations are almost fully randomized in both Pb-Pb and O-O collisions, although the randomization is slightly weaker in O-O.
In other words, the interaction with the glasma is sufficiently strong to wash out the spin polarization generated at very early times.
This observation is relevant not only for the spin-polarization observables by the glasma~\cite{Kumar:2022ylt, Kumar:2023ghs, Yang:2024qpy, Sung:2025gke} but also those by the strong magnetic field in peripheral collisions, e.g., the splitting of global $\Lambda$- and $\bar{\Lambda}$-polarizations, which is so far consistent with zero within experimental uncertainties~\cite{STAR:2023nvo}.
Since the strong magnetic field exists only immediately after the collision~\cite{Hattori:2016emy}, any charge-dependent spin polarization generated at such early times may subsequently be smeared out during the glasma evolution.

The spin polarization also exhibits sizeable local event-by-event fluctuations in the transverse plane.
This is illustrated in the right-most panel of Fig.~\ref{fig:chargedistro}, where the beam-direction component of the rest-frame spin, $s_z$, is shown.
The event-averaged mean polarization approaches zero because the orientations of the glasma fields fluctuate from event to event.
We characterize the magnitude of the fluctuations by the root-mean-square value,
\begin{align}
   \sqrt{\langle\mathcal{P}_z^2\rangle} := \sqrt{   \left\langle \left(    \frac{1}{N}\sum_{n=1}^{N} s_z^n \right)^2    \right\rangle } \;.
\end{align}
From an ensemble of 200 WAGASHI events, we obtain $\sqrt{\langle\mathcal{P}_z^2\rangle_{\rm Pb}}=0.014$ and $\sqrt{\langle\mathcal{P}_z^2\rangle_{\rm O}}=0.085$.  
Interestingly, the Pb-Pb fluctuation is comparable in magnitude to the Glauber-model estimate $\sqrt{\langle\mathcal{P}_z^2\rangle} \approx 0.02$~\cite{giacalone_initial-state-driven_2025}, despite their different physical origins: stochastic interactions with the fluctuating glasma fields in WAGASHI and fluctuations of the initial nucleon configurations in the Glauber model.
The substantially larger value in O-O reflects the stronger finite-number fluctuations in the smaller system.

\paragraph{Conclusions.---}
We have developed a new code, WAGASHI, to study the spacetime evolution of quarks produced via the non-perturbative Schwinger pair production during the earliest stage of the glasma, $\tau \lesssim 0.5\;{\rm fm}/c$.  
We have demonstrated that the glasma strongly reshapes Schwinger-produced quarks by broadening their momentum distributions, erasing their initial spin memory, and generating sizeable charge fluctuations, particularly in small O-O collisions.
The quark charge and spin distributions obtained with WAGASHI provide the initial-state inputs required for the subsequent hydrodynamic evolution of the QGP, including recent spin-hydrodynamic simulations~\cite{matsuda_achieving_2025,singh_spin_2025}.

\paragraph{Acknowledgments---}
Numerical computation in this work was carried out at the Yukawa Institute Computer Facility.
Generative AI (Claude Science) was used to review and find bugs in some code used in this letter.
All suggestions from the AI were carefully reviewed by the authors, and all the code was written by the authors.
Perceptually uniform color maps are used in this study to prevent visual distortion of the data and to better serve those with color vision deficiency~\cite{crameri_misuse_2020}.
This work is supported by JSPS KAKENHI Grant Nos. 24K17058 (H.T.) and 20H00156, JP20H11581, and JP2500449 (C.N.); the RIKEN TRIP initiative (RIKEN Quantum) (H.T.); and the World Premier International Research Center Initiative (WPI) under MEXT, Japan (C.N.).
        
\bibliography{WAGASHI_Letter.bib}

@article{aamodt_centrality_2011,
  title         = {Centrality {Dependence} of the {Charged}-{Particle} {Multiplicity} {Density} at {Midrapidity} in {Pb}-{Pb} {Collisions} at s {NN} = 2.76 {TeV}},
  volume        = {106},
  issn          = {0031-9007, 1079-7114},
  url           = {https://link.aps.org/doi/10.1103/PhysRevLett.106.032301},
  doi           = {10.1103/PhysRevLett.106.032301},
  number        = {3},
  urldate       = {2026-07-08},
  journal       = {Physical Review Letters},
  author        = {{ALICE Collaboration}},
  month         = jan,
  year          = {2011},
  pages         = {032301},
  eprint        = {1012.1657},
  archiveprefix = {arXiv},
  primaryclass  = {nucl-ex}
}

@article{adam_centrality_2016,
  title         = {Centrality {Dependence} of the {Charged}-{Particle} {Multiplicity} {Density} at {Midrapidity} in {Pb}-{Pb} {Collisions} at s {NN} = 5.02 {TeV}},
  volume        = {116},
  issn          = {0031-9007, 1079-7114},
  url           = {https://link.aps.org/doi/10.1103/PhysRevLett.116.222302},
  doi           = {10.1103/PhysRevLett.116.222302},
  number        = {22},
  urldate       = {2026-02-06},
  journal       = {Physical Review Letters},
  author        = {{ALICE Collaboration}},
  month         = jun,
  year          = {2016},
  pages         = {222302},
  eprint        = {1512.06104},
  archiveprefix = {arXiv},
  primaryclass  = {nucl-ex}
}

@article{Aleksandrov:2018zso,
  author        = {Aleksandrov, I. A. and Plunien, G. and Shabaev, V. M.},
  title         = {{Locally-constant field approximation in studies of electron-positron pair production in strong external fields}},
  eprint        = {1811.01419},
  archiveprefix = {arXiv},
  primaryclass  = {hep-ph},
  doi           = {10.1103/PhysRevD.99.016020},
  journal       = {Phys. Rev. D},
  volume        = {99},
  number        = {1},
  pages         = {016020},
  year          = {2019}
}

@misc{alice_centrality_2025,
  title         = {Centrality dependence of charged-particle pseudorapidity density at midrapidity in {Pb}-{Pb} collisions at s {NN} = 5.36 {TeV}},
  urldate       = {2025-04-07},
  author        = {{ALICE Collaboration}},
  month         = apr,
  year          = {2025},
  eprint        = {2504.02505},
  archiveprefix = {arXiv},
  primaryclass  = {nucl-ex}
}

@misc{ALICE:2026zck,
  author        = {Ali Hassan Abdallah, Dana and others},
  collaboration = {ALICE},
  title         = {{Evidence for parton energy loss in oxygen$-$oxygen collisions at $\mathbf{\sqrt{s_{\rm NN}}=5.36}$ TeV}},
  eprint        = {2606.19967},
  archiveprefix = {arXiv},
  primaryclass  = {nucl-ex},
  reportnumber  = {CERN-EP-2026-171},
  month         = {6},
  year          = {2026}
}

@article{avramescu_simulating_2023,
  title         = {Simulating jets and heavy quarks in the {Glasma} using the colored particle-in-cell method},
  volume        = {107},
  issn          = {2470-0010, 2470-0029},
  doi           = {10.1103/PhysRevD.107.114021},
  number        = {11},
  urldate       = {2024-10-03},
  journal       = {Physical Review D},
  author        = {Avramescu, Dana and Băran, Virgil and Greco, Vincenzo and Ipp, Andreas and Müller, David I. and Ruggieri, Marco},
  month         = jun,
  year          = {2023},
  pages         = {114021},
  eprint        = {2303.05599},
  archiveprefix = {arXiv},
  primaryclass  = {hep-ph}
}

@article{Baier:2000sb,
  author        = {Baier, R. and Mueller, Alfred H. and Schiff, D. and Son, D. T.},
  title         = {{'Bottom up' thermalization in heavy ion collisions}},
  eprint        = {hep-ph/0009237},
  archiveprefix = {arXiv},
  doi           = {10.1016/S0370-2693(01)00191-5},
  journal       = {Phys. Lett. B},
  volume        = {502},
  pages         = {51--58},
  year          = {2001}
}

@article{Banerjee:1989by,
  author       = {Banerjee, B. and Bhalerao, R. S. and Ravishankar, V.},
  title        = {{Equilibration of the Quark - Gluon Plasma Produced in Relativistic Heavy Ion Collisions}},
  reportnumber = {TIFR/TH/89-2},
  doi          = {10.1016/0370-2693(89)91041-1},
  journal      = {Phys. Lett. B},
  volume       = {224},
  pages        = {16--20},
  year         = {1989}
}

@article{Bargmann:1959gz,
  author  = {Bargmann, V. and Michel, Louis and Telegdi, V. L.},
  editor  = {Damour, Thibault and Todorov, Ivan and Zhilinskii, Boris},
  title   = {{Precession of the polarization of particles moving in a homogeneous electromagnetic field}},
  doi     = {10.1103/PhysRevLett.2.435},
  journal = {Phys. Rev. Lett.},
  volume  = {2},
  pages   = {435--436},
  year    = {1959}
}

@article{barrera_cabodevila_quark_2025,
  title   = {Quark production in the bottom-up thermalization},
  volume  = {871},
  issn    = {03702693},
  url     = {https://linkinghub.elsevier.com/retrieve/pii/S0370269325007452},
  doi     = {10.1016/j.physletb.2025.139987},
  urldate = {2025-12-24},
  journal = {Physics Letters B},
  author  = {Barrera Cabodevila, Sergio and Du, Xiaojian and Salgado, Carlos A. and Wu, Bin},
  month   = dec,
  year    = {2025},
  pages   = {139987}
}

@article{Batalin:1976uv,
  author       = {Batalin, I. A. and Matinyan, Sergei G. and Savvidy, G. K.},
  title        = {{Vacuum Polarization by a Source-Free Gauge Field}},
  reportnumber = {EFI-198-44-76-YEREVAN},
  journal      = {Sov. J. Nucl. Phys.},
  volume       = {26},
  pages        = {214},
  year         = {1977}
}

@article{Berges:2020fwq,
  author        = {Berges, J{\"u}rgen and Heller, Michal P. and Mazeliauskas, Aleksas and Venugopalan, Raju},
  title         = {{QCD thermalization: Ab initio approaches and interdisciplinary connections}},
  eprint        = {2005.12299},
  archiveprefix = {arXiv},
  primaryclass  = {hep-th},
  reportnumber  = {CERN-TH-2020-080},
  doi           = {10.1103/RevModPhys.93.035003},
  journal       = {Rev. Mod. Phys.},
  volume        = {93},
  number        = {3},
  pages         = {035003},
  year          = {2021}
}

@article{Bialas:1986mt,
  author       = {Bialas, A. and Czyz, W.},
  title        = {{Production and Collective Motion of $q \bar{q}$ Plasma in Heavy Ion Collisions}},
  reportnumber = {IFJ-1307/PH},
  journal      = {Acta Phys. Polon. B},
  volume       = {17},
  pages        = {635},
  year         = {1986}
}

@article{Bialas:1987en,
  author       = {Bialas, A. and Czyz, W. and Dyrek, A. and Florkowski, W.},
  title        = {{Oscillations of Quark - Gluon Plasma Generated in Strong Color Fields}},
  reportnumber = {TPJU-3/87},
  doi          = {10.1016/0550-3213(88)90035-1},
  journal      = {Nucl. Phys. B},
  volume       = {296},
  pages        = {611--624},
  year         = {1988}
}

@inproceedings{Brewer:2021kiv,
  author        = {Brewer, Jasmine and Mazeliauskas, Aleksas and van der Schee, Wilke},
  title         = {{Opportunities of OO and $p$O collisions at the LHC}},
  booktitle     = {{Opportunities of OO and pO collisions at the LHC}},
  eprint        = {2103.01939},
  archiveprefix = {arXiv},
  primaryclass  = {hep-ph},
  reportnumber  = {CERN-TH-2021-028},
  month         = {3},
  year          = {2021}
}

@article{brown_vacuum_1979,
  title   = {Vacuum polarization in uniform non-{Abelian} gauge fields},
  volume  = {157},
  issn    = {05503213},
  url     = {https://linkinghub.elsevier.com/retrieve/pii/055032137990508X},
  doi     = {10.1016/0550-3213(79)90508-X},
  number  = {2},
  urldate = {2025-12-24},
  journal = {Nuclear Physics B},
  author  = {Brown, Lowell S. and Weisberger, William L.},
  month   = sep,
  year    = {1979},
  pages   = {285--326}
}

@article{Bulanov:2004de,
  author        = {Bulanov, S. S. and Narozhny, N. B. and Mur, V. D. and Popov, V. S.},
  title         = {{On e+ e- pair production by a focused laser pulse in vacuum}},
  eprint        = {hep-ph/0403163},
  archiveprefix = {arXiv},
  doi           = {10.1016/j.physleta.2004.07.013},
  journal       = {Phys. Lett. A},
  volume        = {330},
  pages         = {1--6},
  year          = {2004}
}

@article{Busza:2018rrf,
  author        = {Busza, Wit and Rajagopal, Krishna and van der Schee, Wilke},
  title         = {{Heavy Ion Collisions: The Big Picture, and the Big Questions}},
  eprint        = {1802.04801},
  archiveprefix = {arXiv},
  primaryclass  = {hep-ph},
  reportnumber  = {MIT-CTP-4892, MIT-CTP/4892},
  doi           = {10.1146/annurev-nucl-101917-020852},
  journal       = {Ann. Rev. Nucl. Part. Sci.},
  volume        = {68},
  pages         = {339--376},
  year          = {2018}
}

@article{Cabodevila:2023htm,
  author        = {Cabodevila, Sergio Barrera and Salgado, Carlos A. and Wu, Bin},
  title         = {{Quark production and thermalization of the quark-gluon plasma}},
  eprint        = {2311.07450},
  archiveprefix = {arXiv},
  primaryclass  = {hep-ph},
  doi           = {10.1007/JHEP06(2024)145},
  journal       = {JHEP},
  volume        = {06},
  pages         = {145},
  year          = {2024}
}

@misc{carzon_toward_2021,
  title      = {Toward {Initial} {Conditions} of {Conserved} {Charges} {Part} {II}: {The} {ICCING} {Monte} {Carlo} {Algorithm}},
  shorttitle = {Toward {Initial} {Conditions} of {Conserved} {Charges} {Part} {II}},
  url        = {http://arxiv.org/abs/1911.12454},
  doi        = {10.48550/arXiv.1911.12454},
  urldate    = {2025-09-12},
  publisher  = {arXiv},
  author     = {Carzon, Patrick and Martinez, Mauricio and Sievert, Matthew D. and Wertepny, Douglas E. and Noronha-Hostler, Jacquelyn},
  month      = aug,
  year       = {2021},
  note       = {arXiv:1911.12454 [nucl-th]}
}

@article{CMS:2025bta,
  author        = {Hayrapetyan, Aram and others},
  collaboration = {CMS},
  title         = {{Observation of Suppressed Charged-Particle Production in Ultrarelativistic Oxygen-Oxygen Collisions}},
  eprint        = {2510.09864},
  archiveprefix = {arXiv},
  primaryclass  = {nucl-ex},
  reportnumber  = {CMS-HIN-25-008, CERN-EP-2025-226},
  doi           = {10.1103/89sf-9t1x},
  journal       = {Phys. Rev. Lett.},
  volume        = {136},
  number        = {16},
  pages         = {162301},
  year          = {2026}
}

@misc{CMS:2025tga,
  author        = {Hayrapetyan, Aram and others},
  collaboration = {CMS},
  title         = {{Observation of long-range collective flow in OO and NeNe collisions and implications for nuclear structure studies}},
  eprint        = {2510.02580},
  archiveprefix = {arXiv},
  primaryclass  = {nucl-ex},
  reportnumber  = {CMS-HIN-25-009, CERN-EP-2025-222},
  month         = {10},
  year          = {2025}
}

@misc{collaboration_centrality_2026,
  title     = {Centrality dependence of charged-hadron pseudorapidity distributions in oxygen-oxygen collisions at \${\textbackslash}sqrt\{s\_{\textbackslash}mathrm\{{NN}\}\}\$ = 5.36 {TeV}},
  url       = {http://arxiv.org/abs/2606.02285},
  doi       = {10.48550/arXiv.2606.02285},
  urldate   = {2026-07-31},
  publisher = {arXiv},
  author    = {Collaboration, C. M. S.},
  month     = jun,
  year      = {2026},
  note      = {arXiv:2606.02285 [nucl-ex]}
}

@article{crameri_misuse_2020,
  title   = {The misuse of colour in science communication},
  volume  = {11},
  issn    = {2041-1723},
  url     = {https://www.nature.com/articles/s41467-020-19160-7},
  doi     = {10.1038/s41467-020-19160-7},
  number  = {1},
  urldate = {2024-12-09},
  journal = {Nature Communications},
  author  = {Crameri, Fabio and Shephard, Grace E. and Heron, Philip J.},
  month   = {oct},
  year    = {2020},
  pages   = {5444}
}

@article{Du:2020zqg,
  author        = {Du, Xiaojian and Schlichting, S{\"o}ren},
  title         = {{Equilibration of the Quark-Gluon Plasma at Finite Net-Baryon Density in QCD Kinetic Theory}},
  eprint        = {2012.09068},
  archiveprefix = {arXiv},
  primaryclass  = {hep-ph},
  doi           = {10.1103/PhysRevLett.127.122301},
  journal       = {Phys. Rev. Lett.},
  volume        = {127},
  number        = {12},
  pages         = {122301},
  year          = {2021}
}

@article{dumitru_numerical_2006,
  title     = {Numerical simulation of non-{Abelian} particle-field dynamics},
  volume    = {29},
  copyright = {http://www.springer.com/tdm},
  issn      = {1434-6001, 1434-601X},
  url       = {http://link.springer.com/10.1140/epja/i2005-10300-3},
  doi       = {10.1140/epja/i2005-10300-3},
  number    = {1},
  urldate   = {2025-08-13},
  journal   = {The European Physical Journal A},
  author    = {Dumitru, A. and Nara, Y.},
  month     = jul,
  year      = {2006},
  pages     = {65--69}
}

@article{elze_quark-gluon_1989,
  title    = {Quark-{Gluon} {Transport} {Theory}},
  volume   = {183},
  number   = {3},
  journal  = {Physics Reports},
  author   = {Elze, Hans-Thomas and Heinz, Ulrich},
  year     = {1989},
  pages    = {81--135}
}

@article{Fedotov:2022ely,
  author        = {Fedotov, A. and Ilderton, A. and Karbstein, F. and King, B. and Seipt, D. and Taya, H. and Torgrimsson, G.},
  title         = {{Advances in QED with intense background fields}},
  eprint        = {2203.00019},
  archiveprefix = {arXiv},
  primaryclass  = {hep-ph},
  reportnumber  = {RIKEN-iTHEMS-Report-22},
  doi           = {10.1016/j.physrep.2023.01.003},
  journal       = {Phys. Rept.},
  volume        = {1010},
  pages         = {1--138},
  year          = {2023}
}

@article{Gatoff:1987uf,
  author       = {Gatoff, G. and Kerman, A. K. and Matsui, T.},
  title        = {{The Flux Tube Model for Ultrarelativistic Heavy Ion Collisions: Electrohydrodynamics of a Quark Gluon Plasma}},
  reportnumber = {MIT-CTP-1338},
  doi          = {10.1103/PhysRevD.36.114},
  journal      = {Phys. Rev. D},
  volume       = {36},
  pages        = {114},
  year         = {1987}
}

@article{Gelfand:2016yho,
  author        = {Gelfand, Daniil and Ipp, Andreas and M{\"u}ller, David},
  title         = {{Simulating collisions of thick nuclei in the color glass condensate framework}},
  eprint        = {1605.07184},
  archiveprefix = {arXiv},
  primaryclass  = {hep-ph},
  reportnumber  = {TUW-16-14},
  doi           = {10.1103/PhysRevD.94.014020},
  journal       = {Phys. Rev. D},
  volume        = {94},
  number        = {1},
  pages         = {014020},
  year          = {2016}
}

@article{gelis_chemical_2006,
  title         = {Chemical {Thermalization} in {Relativistic} {Heavy} {Ion} {Collisions}},
  volume        = {96},
  copyright     = {http://link.aps.org/licenses/aps-default-license},
  issn          = {0031-9007, 1079-7114},
  url           = {https://link.aps.org/doi/10.1103/PhysRevLett.96.032304},
  doi           = {10.1103/PhysRevLett.96.032304},
  number        = {3},
  urldate       = {2025-02-14},
  journal       = {Physical Review Letters},
  author        = {Gelis, F. and Kajantie, K. and Lappi, T.},
  month         = jan,
  year          = {2006},
  pages         = {032304},
  eprint        = {hep-ph/0508229},
  archiveprefix = {arXiv},
  primaryclass  = {hep-ph}
}

@article{gelis_quark-antiquark_2005,
  title      = {Quark-antiquark production from classical fields in heavy-ion collisions: 1 + 1 dimensions},
  volume     = {71},
  copyright  = {http://link.aps.org/licenses/aps-default-license},
  issn       = {0556-2813, 1089-490X},
  shorttitle = {Quark-antiquark production from classical fields in heavy-ion collisions},
  url        = {https://link.aps.org/doi/10.1103/PhysRevC.71.024904},
  doi        = {10.1103/PhysRevC.71.024904},
  number     = {2},
  urldate    = {2026-07-29},
  journal    = {Physical Review C},
  author     = {Gelis, F. and Kajantie, K. and Lappi, T.},
  month      = feb,
  year       = {2005},
  pages      = {024904}
}

@article{Gelis:2015kya,
  author        = {Gelis, Francois and Tanji, Naoto},
  title         = {{Schwinger mechanism revisited}},
  eprint        = {1510.05451},
  archiveprefix = {arXiv},
  primaryclass  = {hep-ph},
  doi           = {10.1016/j.ppnp.2015.11.001},
  journal       = {Prog. Part. Nucl. Phys.},
  volume        = {87},
  pages         = {1--49},
  year          = {2016}
}

@misc{giacalone_initial-state-driven_2025,
  title     = {Initial-state-driven spin correlations in high-energy nuclear collisions},
  url       = {http://arxiv.org/abs/2502.13102},
  doi       = {10.48550/arXiv.2502.13102},
  urldate   = {2025-09-12},
  publisher = {arXiv},
  author    = {Giacalone, Giuliano and Speranza, Enrico},
  month     = feb,
  year      = {2025},
  note      = {arXiv:2502.13102 [nucl-th]}
}

@article{Gyulassy:1985oqt,
  author       = {Gyulassy, M. and Iwazaki, A.},
  title        = {{QUARK AND GLUON PAIR PRODUCTION IN SU(N) COVARIANT CONSTANT FIELDS}},
  reportnumber = {LBL-20318},
  doi          = {10.1016/0370-2693(85)90711-7},
  journal      = {Phys. Lett. B},
  volume       = {165},
  pages        = {157--161},
  year         = {1985}
}

@article{Hattori:2016emy,
  author        = {Hattori, Koichi and Huang, Xu-Guang},
  title         = {{Novel quantum phenomena induced by strong magnetic fields in heavy-ion collisions}},
  eprint        = {1609.00747},
  archiveprefix = {arXiv},
  primaryclass  = {nucl-th},
  reportnumber  = {RBRC-1202},
  doi           = {10.1007/s41365-016-0178-3},
  journal       = {Nucl. Sci. Tech.},
  volume        = {28},
  number        = {2},
  pages         = {26},
  year          = {2017}
}

@article{Hidaka:2022dmn,
  author        = {Hidaka, Yoshimasa and Pu, Shi and Wang, Qun and Yang, Di-Lun},
  title         = {{Foundations and applications of quantum kinetic theory}},
  eprint        = {2201.07644},
  archiveprefix = {arXiv},
  primaryclass  = {hep-ph},
  reportnumber  = {KEK-TH-2390, J-PARC-TH-0267, RIKEN-iTHEMS-Report-22},
  doi           = {10.1016/j.ppnp.2022.103989},
  journal       = {Prog. Part. Nucl. Phys.},
  volume        = {127},
  pages         = {103989},
  year          = {2022}
}

@article{Kajantie:1985jh,
  author       = {Kajantie, K. and Matsui, T.},
  title        = {{Decay of Strong Color Electric Field and Thermalization in Ultrarelativistic Nucleus-Nucleus Collisions}},
  reportnumber = {HU-TFT-85-36},
  doi          = {10.1016/0370-2693(85)90343-0},
  journal      = {Phys. Lett. B},
  volume       = {164},
  pages        = {373--378},
  year         = {1985}
}

@article{Kumar:2022ylt,
  author        = {Kumar, Avdhesh and M{\"u}ller, Berndt and Yang, Di-Lun},
  title         = {{Spin polarization and correlation of quarks from the glasma}},
  eprint        = {2212.13354},
  archiveprefix = {arXiv},
  primaryclass  = {nucl-th},
  doi           = {10.1103/PhysRevD.107.076025},
  journal       = {Phys. Rev. D},
  volume        = {107},
  number        = {7},
  pages         = {076025},
  year          = {2023}
}

@article{Kumar:2023ghs,
  author        = {Kumar, Avdhesh and M{\"u}ller, Berndt and Yang, Di-Lun},
  title         = {{Spin alignment of vector mesons by glasma fields}},
  eprint        = {2304.04181},
  archiveprefix = {arXiv},
  primaryclass  = {nucl-th},
  doi           = {10.1103/PhysRevD.108.016020},
  journal       = {Phys. Rev. D},
  volume        = {108},
  number        = {1},
  pages         = {016020},
  year          = {2023}
}

@article{Kurkela:2018oqw,
  author        = {Kurkela, Aleksi and Mazeliauskas, Aleksas},
  title         = {{Chemical equilibration in weakly coupled QCD}},
  eprint        = {1811.03068},
  archiveprefix = {arXiv},
  primaryclass  = {hep-ph},
  reportnumber  = {CERN-TH-2018-239},
  doi           = {10.1103/PhysRevD.99.054018},
  journal       = {Phys. Rev. D},
  volume        = {99},
  number        = {5},
  pages         = {054018},
  year          = {2019}
}

@article{Lappi:2006fp,
  author        = {Lappi, T. and McLerran, L.},
  title         = {{Some features of the glasma}},
  eprint        = {hep-ph/0602189},
  archiveprefix = {arXiv},
  reportnumber  = {BNL-NT-06-10},
  doi           = {10.1016/j.nuclphysa.2006.04.001},
  journal       = {Nucl. Phys. A},
  volume        = {772},
  pages         = {200--212},
  year          = {2006}
}

@article{loizides_glauber_2026,
  title         = {Glauber predictions for oxygen and neon collisions at energies available at the {CERN} {Large} {Hadron} {Collider}},
  volume        = {113},
  issn          = {2469-9985, 2469-9993},
  url           = {https://link.aps.org/doi/10.1103/mkp8-zgxh},
  doi           = {10.1103/mkp8-zgxh},
  number        = {1},
  urldate       = {2026-07-31},
  journal       = {Physical Review C},
  author        = {Loizides, Constantin},
  month         = jan,
  year          = {2026},
  pages         = {014914},
  eprint        = {2507.05853},
  archiveprefix = {arXiv},
  primaryclass  = {nucl-th}
}

@misc{matsuda_achieving_2025,
  title     = {Achieving angular-momentum conservation with physics-informed neural networks in computational relativistic spin hydrodynamics},
  url       = {http://arxiv.org/abs/2512.17971},
  doi       = {10.48550/arXiv.2512.17971},
  urldate   = {2025-12-24},
  publisher = {arXiv},
  author    = {Matsuda, Hidefumi and Hattori, Koichi and Murase, Koichi},
  month     = dec,
  year      = {2025}
}

@article{Matsuda:2024mmr,
  author        = {Matsuda, Hidefumi and Huang, Xu-Guang},
  title         = {{Simulation of a (3+1)D glasma in Milne coordinates: Topological charge, eccentricity, and angular momentum}},
  eprint        = {2409.08742},
  archiveprefix = {arXiv},
  primaryclass  = {hep-ph},
  doi           = {10.1103/PhysRevD.110.114032},
  journal       = {Phys. Rev. D},
  volume        = {110},
  number        = {11},
  pages         = {114032},
  year          = {2024}
}

@article{McDonald:2023qwc,
  author        = {McDonald, Scott and Jeon, Sangyong and Gale, Charles},
  title         = {{3+1D initialization and evolution of the glasma}},
  eprint        = {2306.04896},
  archiveprefix = {arXiv},
  primaryclass  = {hep-ph},
  doi           = {10.1103/PhysRevC.108.064910},
  journal       = {Phys. Rev. C},
  volume        = {108},
  number        = {6},
  pages         = {064910},
  year          = {2023}
}

@article{McLerran:1993ka,
  author        = {McLerran, Larry D. and Venugopalan, Raju},
  title         = {{Gluon distribution functions for very large nuclei at small transverse momentum}},
  eprint        = {hep-ph/9311205},
  archiveprefix = {arXiv},
  reportnumber  = {TPI-MINN-93-52-T, NUC-MINN-93-28-T, UMN-TH-1224-93},
  doi           = {10.1103/PhysRevD.49.3352},
  journal       = {Phys. Rev. D},
  volume        = {49},
  pages         = {3352--3355},
  year          = {1994}
}

@article{McLerran:1993ni,
  author        = {McLerran, Larry D. and Venugopalan, Raju},
  title         = {{Computing quark and gluon distribution functions for very large nuclei}},
  eprint        = {hep-ph/9309289},
  archiveprefix = {arXiv},
  reportnumber  = {TPI-MINN-93-44-T, NUC-MINN-93-24-T, HEP-UMN-TH-1220-93},
  doi           = {10.1103/PhysRevD.49.2233},
  journal       = {Phys. Rev. D},
  volume        = {49},
  pages         = {2233--2241},
  year          = {1994}
}

@article{McLerran:1994vd,
  author        = {McLerran, Larry D. and Venugopalan, Raju},
  title         = {{Green's functions in the color field of a large nucleus}},
  eprint        = {hep-ph/9402335},
  archiveprefix = {arXiv},
  reportnumber  = {TPI-MINN-94-7-T, NUC-MINN-94-2-T, HEP-MINN-94-1242-T},
  doi           = {10.1103/PhysRevD.50.2225},
  journal       = {Phys. Rev. D},
  volume        = {50},
  pages         = {2225--2233},
  year          = {1994}
}

@phdthesis{muller_simulations_2020,
  title         = {Simulations of the {Glasma} in 3+{1D}},
  url           = {http://arxiv.org/abs/1904.04267},
  urldate       = {2024-11-07},
  school        = {Technischen Universitat Wien},
  author        = {Müller, David},
  month         = sep,
  year          = {2020},
  eprint        = {1904.04267},
  archiveprefix = {arXiv},
  primaryclass  = {hep-ph}
}

@article{Nayak:2005pf,
  author        = {Nayak, Gouranga C.},
  title         = {{Non-perturbative quark-antiquark production from a constant chromo-electric field via the Schwinger mechanism}},
  eprint        = {hep-ph/0510052},
  archiveprefix = {arXiv},
  reportnumber  = {YITP-SB-05-33},
  doi           = {10.1103/PhysRevD.72.125010},
  journal       = {Phys. Rev. D},
  volume        = {72},
  pages         = {125010},
  year          = {2005}
}

@article{Nikishov:1969tt,
  author  = {Nikishov, A. I.},
  title   = {{Pair production by a constant external field}},
  journal = {Zh. Eksp. Teor. Fiz.},
  volume  = {57},
  pages   = {1210--1216},
  year    = {1969}
}

@misc{plumberg_bsq_2024,
  title     = {{BSQ} {Conserved} {Charges} in {Relativistic} {Viscous} {Hydrodynamics} solved with {Smoothed} {Particle} {Hydrodynamics}},
  url       = {http://arxiv.org/abs/2405.09648},
  doi       = {10.48550/arXiv.2405.09648},
  urldate   = {2025-04-22},
  publisher = {arXiv},
  author    = {Plumberg, Christopher and Almaalol, Dekrayat and Dore, Travis and Mroczek, Débora and Martín, Jordi Salinas San and Serenone, Willian M. and Spychalla, Lydia and Carzon, Patrick and Sievert, Matthew D. and Gardim, Fernando G. and Noronha-Hostler, Jacquelyn},
  month     = may,
  year      = {2024},
  note      = {arXiv:2405.09648 [nucl-th]}
}

@article{Schwinger:1951nm,
  author  = {Schwinger, Julian S.},
  editor  = {Milton, K. A.},
  title   = {{On gauge invariance and vacuum polarization}},
  doi     = {10.1103/PhysRev.82.664},
  journal = {Phys. Rev.},
  volume  = {82},
  pages   = {664--679},
  year    = {1951}
}

\end{document}